\documentstyle[preprint,aps]{revtex}
\tighten

\begin{document}

\draft

\title {Fingering Instability in Combustion}

\author { O. Zik$^{(1)}$, Z. Olami$^{(2)}$ and E. Moses$^{(1)}$}
 \address{$^{(1)}$Department of Physics of Complex Systems and
$^{(2)}$Department of
Chemical Physics,\\
  The Weizmann Institute of Science, Rehovot 76100, Israel}
 \maketitle

 \begin{abstract}
A thin solid (e.g., paper), burning against an oxidizing wind, develops a
fingering instability with two decoupled length scales.
The spacing between
fingers is determined by the P\'eclet number (ratio between advection and
diffusion). The finger width is determined by the degree two dimensionality. Dense fingers develop by recurrent
tip splitting. The effect is
observed when vertical mass transport
(due to gravity)  is suppressed.
The experimental results quantitatively
verify a model based on diffusion limited
transport.

\vspace{2 pc} 

\end{abstract}

\mediumtext

Harnessing combustion is a basic element of technology, from
lighting a candle \cite {FARADAY} to
propelling a spacecraft.  The transition to unrestrained conflagrations is often determined by fundamental issues
regarding the stability of combustion fronts.  Although characterized
 by tens of concurrent chemical reactions which depend exponentially on temperature \cite {SIVA}, these complex
systems are governed by the availability of three basic constituents: oxidant, fuel and heat. The experimental
control of these
coupled fields can be considerably improved by studying the slow combustion
regime termed smoldering \cite{SMOLD}.  The evolution of a smoldering front
is of particular importance for fire safety, since the initial development
of a slow burning process will determine whether the fire grows or
extinguishes.

The stability of one dimensional fronts has been extensively studied in simpler, less
reactive, growth systems\cite{REV}. 
In these systems the pattern is typically characterized by one length scale (the fastest growing wavelength
in the linear spectrum). Notable examples are solidification\cite{LANGER}, liquid-crystals\cite{SIMON} and
viscous fronts\cite{ST58}.
Interfacial instabilities are also known in combustion. The most studied is the
thermal-diffusion instability in `premixed'  flames (where the oxygen and gaseous fuel are mixed before they
react). This instability results from the competition between the destabilizing effect of molecular transport and
the stabilizing effect of heat transport \cite {SIVA}, \cite{CLAVIN}, \cite {GORMAN}. Previous studies of flame
propagation over thin solid fuels focused mainly on flame spread rate\cite{FREY},\cite{OLSON}. A cellular
structure observed in this system was recently related to the thermal-diffusion instability \cite{RONNEY}.

In this paper we report a new effect: a combustion front in a quasi two dimensional geometry, exhibits a {\it
directional fingering instability} with two  decoupled  length scales (the finger width and the spacing
between fingers).  The non-dimensional control parameter is the P\'eclet number $Pe$ which measures the
relative importance of molecular advection and diffusion \cite{PEC}.  The characteristic scale (finger width) is
effectively independent of $Pe$. It is determined by the degree of two dimensionality. On the
other hand, the distance between fingers is self consistently determined by $Pe$. 
The fingering effect has independently  been observed in a recent
experiment conducted in space \cite{NASA}. 
Our measurements are performed in the experimentally preferable smoldering regime, where the front can be
reproducibly controlled. The oxidizing gas is supplied in a uniform flow, opposite to the direction of the front
propagation (i.e., the fuel and oxygen are fed to the reaction from the same side).  This configuration enhances the
destabilizing effect of reactant transport, which plays the same role as in the thermal-diffusion instability. 
Although heat transport plays a key role in the instability, our phenomenological model shows that reactant
transport alone determines the front velocity and spacing between fingers. The measurements are in remarkable
agreement with this model (no free parameters).


The setup is shown schematically in Fig. ~\ref{Lsetupfig}.  The experiment is initiated by igniting a thin fuel
sample along a line at its downstream edge.  The main
experimental controls are the flow velocity of oxygen
$V_{O2}$ and the vertical gap between plates
$h$. The `fuel' is
filter paper. However, the effect is  fuel independent. It occurs in all the combustible materials that we
examined, e.g., standard stationery paper, cellulose dialysis bag and  polyethylene sheets. In many of these
materials the front involves melting and a glowing flame which complicates the measurements. 



The front is smooth when the oxygen flow velocity $V_{O2}$ is higher than some critical value (Fig.
~\ref{Lpattvsflow}$a$). When $V_{O2}$ is slightly decreased the front develops a structure which marks the
onset of instability (Fig. ~\ref{Lpattvsflow}$b$). Further decreasing $V_{O2}$ yields a periodic cellular
structure or `cusps' (Fig. ~\ref{Lpattvsflow}$c$). As $V_{O2}$ is gradually decreased, the cusps become sharper
until the peaks separate and a fingering pattern is formed. 
A typical fingering pattern is shown in Fig.
~\ref{Lpattvsflow}$d$.  It develops by recurrent tip splitting. Fingers that are closer to the oxygen source
{\it screen} neighboring fingers from the source of oxygen. The screened  fingers stop growing and the tips of the
screening fingers split. The splitting is such that both the average finger width and the spacing between the
fingers remain unchanged. When
$V_{O2}$ is further decreased the spacing between fingers increases and we obtain sparse fingers with no tip
splitting (Fig. ~\ref{Lpattvsflow}$e$). Combustion occurs only in the limited vicinity of the tip, there
is no reaction behind the front (along the fingers). A bright spot (like the one in the central finger
in Fig. ~\ref{Lpattvsflow}$e$) signals a local transition from `smoldering' to `flame', which typically appears just
before a tip stops growing. Fig. ~\ref{Lpattvsflow}$c-e$ shows the existence of two well-defined length scales
in the fingering pattern, the average finger width
$w$ and the average spacing between fingers $d$.  While $w$
depends very weakly on
$V_{O2}$, $d$ is a rapidly decreasing function of it. The maximal value of $d$ is comparable to the
system size $d \sim L$. The minimal value of $d$ (zero) is reached when  the increase in oxygen flow wipes out
the pattern (Fig. ~\ref{Lpattvsflow}$c$). Tip splitting is observed at $d \sim w$. 



 

The above observations are explained in terms of the following phenomenological picture. At high values of
$V_{O2}$ the uniformly fed front is smooth.  As the supply is decreased,
small bumps that exist along  the
interface  begin to compete over the oxygen. This mechanism resemble the  thermal diffusive instability \cite
{SIVA}.  The consumption of oxygen drive lateral diffusion currents. When the oxygen supply is further decreased,
the bumps consume all the oxygen that is available   in their vicinity, and separate into distinct fingers. The
`upstream' fingers (which are closer to the oxygen source) prevail and then tip-split to maintain the same
average $d$.  As the oxygen  supply is further decreased $d$ grows. At a certain stage 
$d$ is sufficiently large to allow fingering without
screening (and tip splitting). Unlike $d$, $w$ is independent of oxygen supply. It depends on heat transport which
is neglected in the  above simplified picture.

The two dominant oxygen transport mechanisms are diffusion and advection. In the quasi $2d$
configuration, the P\'eclet number is 
$Pe={{V_{O2}h}\over D}$\cite{Da}.  Dimensional analysis leads us to expect that the measured quantities $u$, $d$
and $w$ depend on the dimensional
parameters $h$, $V_{O2}$ and $D$ and on $Pe$. 
Thus, we non-dimensionalize  $\tilde w= {w\over h}$, 
 $\tilde d={d\over h}$ and $\tilde u={u\over V_{O2}}$.


In the oxygen limited case expect all the available oxygen to be consumed and write a {\it mass conservation}
equation (see the reference in \cite{SMOLD}):
\begin {equation}
 \tilde u{{\tilde w}\over {\tilde w+\tilde d}} = A
\label {A}
\end{equation}  
where the constant $A$, defined in \cite{REFA}, represents the stoichiometry.
 ${w}\over {w+d}$ is the fraction of burned area, resulting from integration over the width of the sample $L$. 
The deficiency of oxygen produces concentration gradients and consequently {\it lateral diffusion currents} (see
Fig.  ~\ref{LFLOW}.), given by $j_x =D \nabla_x
\rho_{0_2}$, where $\rho_{0_2}$ is the oxygen concentration. The gradient is typically over a distance of $d+w$.
At steady state, the lateral current satisfies
$j_x =u \rho_{02}$ up to a proportionality constant of order unity. 
{
 This yields
 an equation for the velocity of the diffusion limited growth \cite{TRIVEDI} 
\begin {equation}  
{\tilde u= {1\over {(\tilde w +\tilde d)}}{Pe}^{-1}}
\label{u}
\end{equation}  Combining Eq. ~\ref{A} and Eq. ~\ref{u} gives
\begin {eqnarray}
 \tilde d =\sqrt{{{\tilde w}\over{A}}}
 {Pe}^{-{1\over 2}}-\tilde w
\label {d1} \\
  \tilde u=\sqrt{{{A}\over{\tilde w}}} {Pe}^{-{1\over 2}}
\label {u1}\\
\nonumber
\end{eqnarray}

 For $Pe$ greater than some critical value $Pe_{c}$ we observe a  flat front (Fig. ~\ref {Lpattvsflow}).  For $Pe$
smaller than another critical value $Pe_{c1}$ we observe fingers.  We regard the narrow band  $Pe_{c1}\le Pe\le
Pe_{c}$ as onset regime. In this regime we observe a connected  front with a cellular structure. Eq. ~\ref{d1}
yields the critical $Pe_{c1}$ for the onset of a  connected front, when $d$ goes to zero: 
 \begin {equation} Pe_{c1} = {1\over{A \tilde w}}
\label {criticalP}
\end{equation}

The mass conservation equation (Eq. ~\ref{A}) holds for $Pe \le Pe_{c}$. In dimensional units, it predicts a linear
dependence of the quantity of burned material per unit time
${{uw }/{(w+d)}}$ on
$V_{O2}$. This dependence is shown in Fig. ~\ref{Lgammau}. The measured slope is $A=0.043\pm0.005$. The
predicted slope is $A=0.051\pm0.005$  \cite{REFA}. The numerical agreement (within experimental error)
shows that the system  is well within the oxygen deficient regime. The front velocity
$u$ is shown in the insert. Below a critical value corresponding to 
$Pe_{c1}$ (arrow in the figure), it behaves as $u \sim {V_{O2}}^{{1\over 2}}$ as predicted by Eq. ~\ref{u1}.
At $Pe_{c1}$ there is a crossover to the linear dependence predicted by Eq. ~\ref{A} with
$d=0$.



The dependence of the lengths $d$ and $w$ on
$Pe$ is shown in Fig. ~\ref{Ldw}. The ratio between $w$ and $d$ characterizes the pattern. It is
determined by $Pe$. 
Sparse fingers are observed at ${{d}\over {w}}>>1$ ($ 0.1\le Pe \le 2.6$),  tip splitting at $ {{1}\over {4}}\le
{{d}\over{w}} \le {{5}\over {4}}$ ($2.6<Pe\le Pe_{c1}$). Experimentally $Pe_{c1}=17\pm1$. The predicted
value is $Pe_{c1}=1/{Aw}\approx 21$  (Eq. ~\ref{criticalP}). 
 A connected front is observed at $Pe_{c1}\le Pe \le Pe_{c}$. In this regime $d=0$ and $w$ is the characteristic
`cusp' size. Experimentally $Pe_{c}=22 \pm1$ \cite{ONSET}. $w$ and
$d$ are not defined for $Pe\ge Pe_{c}$.  The continuous line in Fig.
~\ref{Ldw} shows the theoretical prediction (Eq. ~\ref{d1}). It is in very good agreement with the data points.
These measurements were performed by varying $V_{O2}$ at $h=0.5 cm$. Other values of $h$ (in the range $0.2
cm\le h\le 1 cm$) show the same qualitative dependence of $d$ on $Pe$. We conjecture that the slow linear
increase of $w$ as a function of $Pe$ is related to the limitation posed by the reaction rate \cite {Da}.


The lengths $w$
and
$d$ are weakly coupled via Eq. ~\ref {d1}. Fig. ~\ref{Lonefinger} exemplifies an experiment in which we isolated
$w$ from
$d$ in two narrow pieces of paper ($L=1 cm$ and $L=0.6 cm$). Both fronts
propagated at the same velocity $u$ and with the same $w$  of the main trunk, as in the wide system (within
experimental scatter).  At $L=1cm$ we observe periodic attempt to tip split which can not be accomplished
(since $L<w+d$). We conclude that the instability is not a collective effect but a feature of the single finger. The
competition over oxygen determines
$d$ but not $w$.


The weak dependence of $w$ on $Pe$ allows us to measure
$w$ as a function of
$h$ in a wide range of $Pe$. Fig. ~\ref{LWH} shows that $w$ is dominated by the degree of
two-dimensionality $h$ \cite{ONSET}. We conjecture that the origin of this dependence is related to
heat transport, via convection of combustion products. This conjecture is consistent with our observation that
$w$ is a decreasing function of the fraction of cooling ($N_2$) gas  in the feeding mixture.  The linear
dependence of the characteristic length of the instability on the spacing between plates is known to occur in this
 (`Hele Shaw') geometry in viscous fronts\cite {REV}.  To some extent,
$w$ also depends on the features of the fuel, heat conductivity of the bottom plate and  the relative nitrogen
concentration in the feeding gas.


In conclusion, the complex burning process exhibits a rich but controllable
pattern.  A flat piece of fuel  (e.g. paper), forced to burn against an
oxidizing wind, develops a steady fingering
state.  This new effect occurs when vertical flow is suppressed.
There are two key processes, transport of heat and transport of reactants,
and two effectively
decoupled length scales: the finger width and the spacing between fingers.
The characteristic length (finger width) is very weakly dependent
on the control parameter (P\'eclet number). It is determined by the
geometry (degree of two dimensionality). The spacing between
fingers is
determined by the P\'eclet number. This dependence is in very good
quantitative agreement with a
phenomenological model which we base on reactant transport. Significant
effort must be devoted to the onset
regime, where preliminary observations already revealed interesting
time-dependent modes.
\clearpage

%

\begin{figure}  

\caption{Schematic representation of the setup. 1. Transparent top, 2. 
variable gap between top and bottom plates ($h$)
 3. outflow of combustion products, 4. spacers to control $h$, 5.  ignition wire
(tungsten), 6. heat conducting boundaries (also serves to decrease
the supply of oxygen from the sides of the sample),   7. flame front, 8. fuel (Whatman filter paper
$\# 2$, $20\times 20cm^2$), 9. interchangeable bottom plate (to study of the role of the boundary's heat
conductivity) 10. uniform  flow of $O_2/N_2$ (the uniformity was verified with  smoke
visualization), 11. gas diffuser, 12. gas inlet. }
\label{Lsetupfig}

\caption{ The instability as a function of
oxygen flow $V_{O2}$ (decreasing from $a$ to $e$).
$(a)$ smooth front, $(b)$ irregular front, $(c)$ periodic pattern, $(d)$ fingering pattern with tip
splitting, $(e)$ fingering without tip splitting.  
The pattern defines two lengths the finger width $w$ and spacing between fingers $d$. 
Both lengths stabilize within a short initial transient.
The scale bars are
$1 cm$. The gap between plates is $h=0.5 cm$. The oxygen flow velocity $V_{O2}$ is directed downwards and the
front velocity 
$u$ is directed upwards. The values are (top to bottom):
$V_{O2} =11.4, 10.2, 9.2 ,1.3, 0.1 cm/s$ and  $u=  0.5, 0.48, 0.41, 0.14, 0.035,\pm0.01 cm/s$.  }
\label{Lpattvsflow}

\caption{The schematic flow field. The
flow lines are diverted by lateral diffusion which is driven by the deficiency of oxygen at
the tips.}
\label{LFLOW}

\caption {Mass conservation
, as 
predicted by  
Eq. 1
The insert shows $u$ as a
function of $V_{O2}$. The arrow corresponds to  $Pe_{c1}$.
We verified that Eq. 1
 is also satisfied when the  front propagates with the
wind and when the parameter being varied is the fraction of oxygen in
the feeding gas ($0.5 \le a\le 1$). 
The  oxygen molecular  diffusion coefficient is taken to be  $D=0.25 cm^2/s$.
 We neglect its temperature
dependence.}
\label{Lgammau}

\caption{The length scales $d$ (full circles) and 
$w$  (empty circles) as a function of
$Pe$. $d$ is determined by the  driving  parameter, while $w$ is  weakly
influenced by it. The slope of
$w(Pe)$ is 
$0.01\pm 0.002 cm$ (dashed line). The continuous line  is a plot of the  RHS of 
 It fits
the measured $d$ with no adjustable parameter. }
\label{Ldw}

\caption{The instability of an isolated finger ($w=0.5cm$) is manifested as a periodic  attempt to
tip split in a thin paper of width $L=1cm$ 
(bottom). It does not occur at $L=0.6cm$. The
distance between the two fingers was
$10cm$, ensuring no interaction between them ($h=0.5 cm$, $Pe=6$). }
\label{Lonefinger}

\caption{The degree of two dimensionallity ($h$) determines $w$ in a linear
fashion. The slope is $1.78 \pm 0.4$. The data points are averages over a range of $V_{O2}$ values. The spread is
represented by the error bars.  The empty circles at $h=0.5 cm$ are the  data
points of Fig.  }
\label{LWH}

 \end {figure}
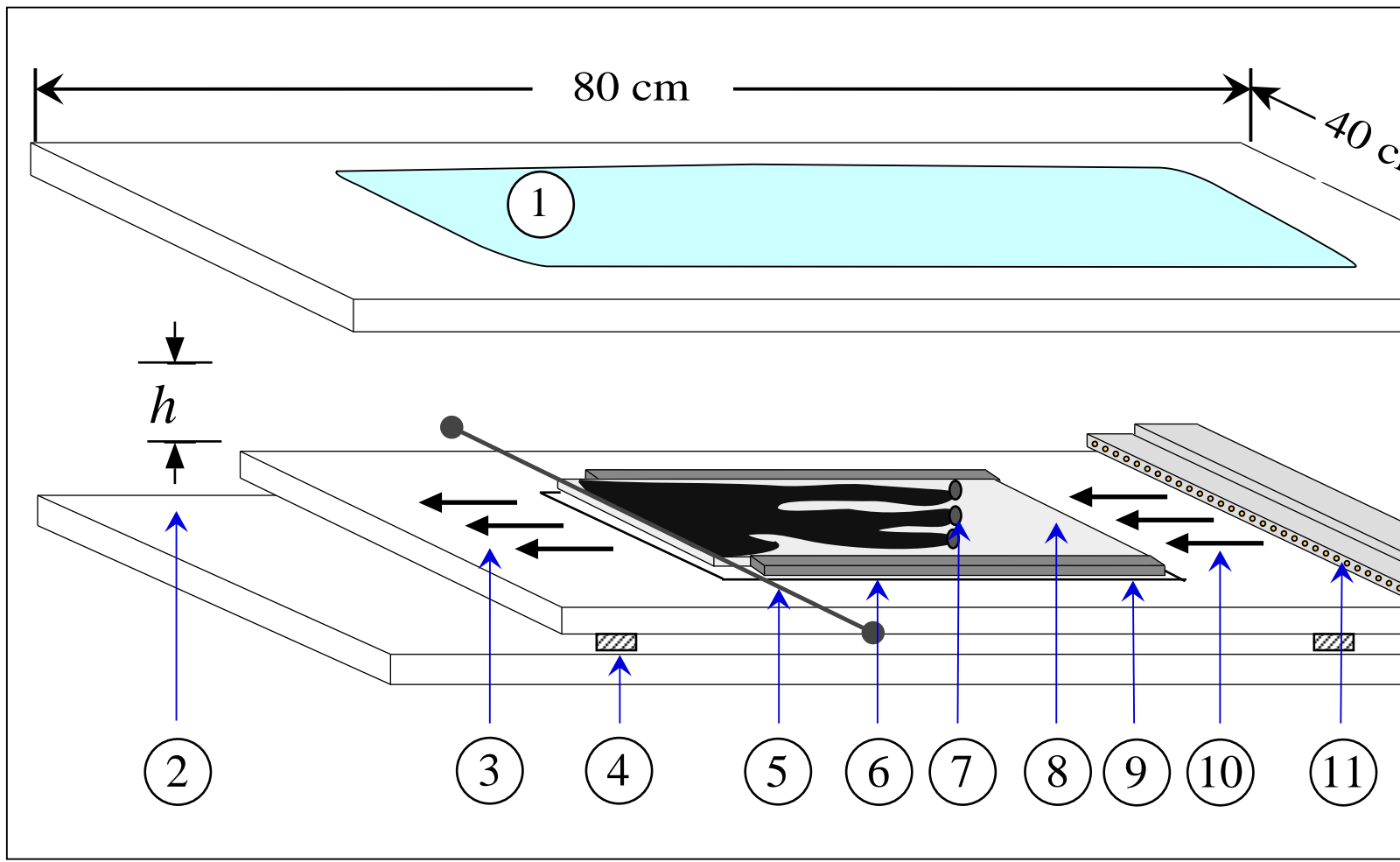
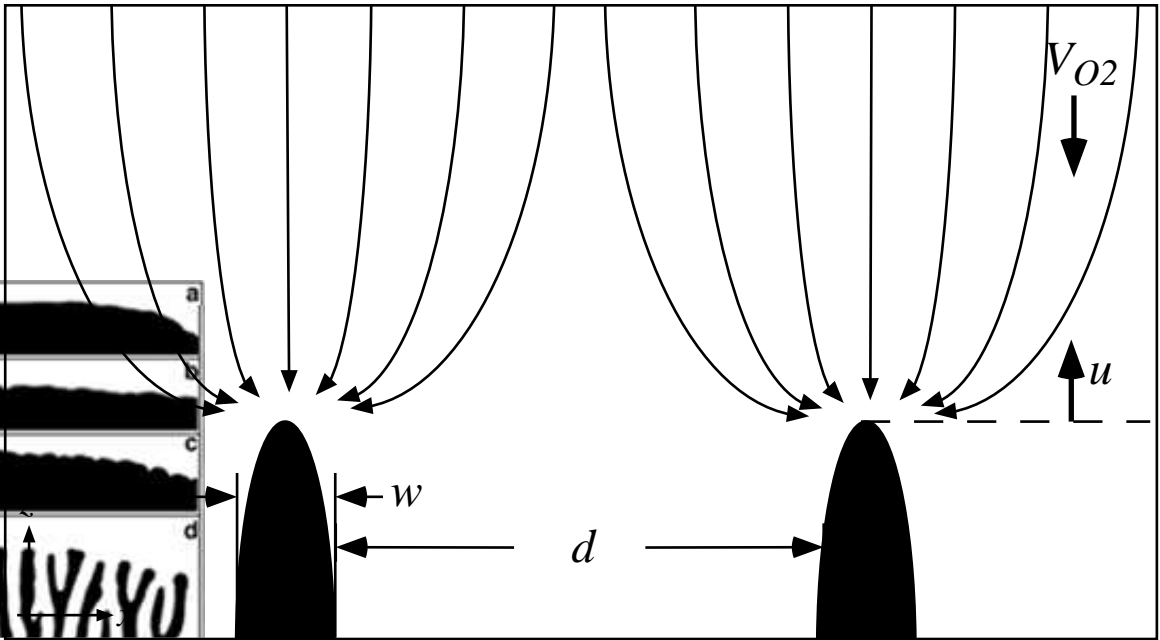
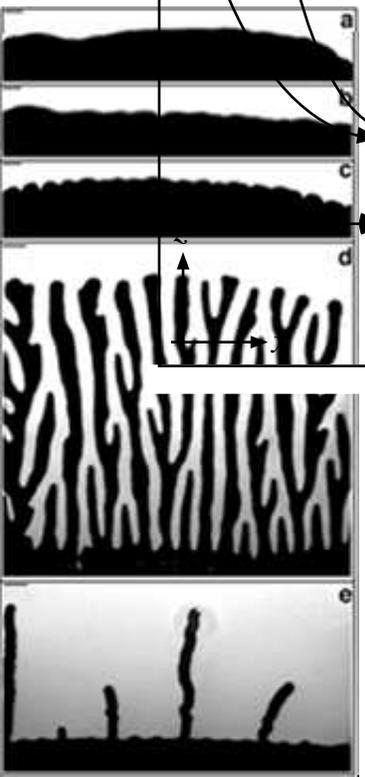
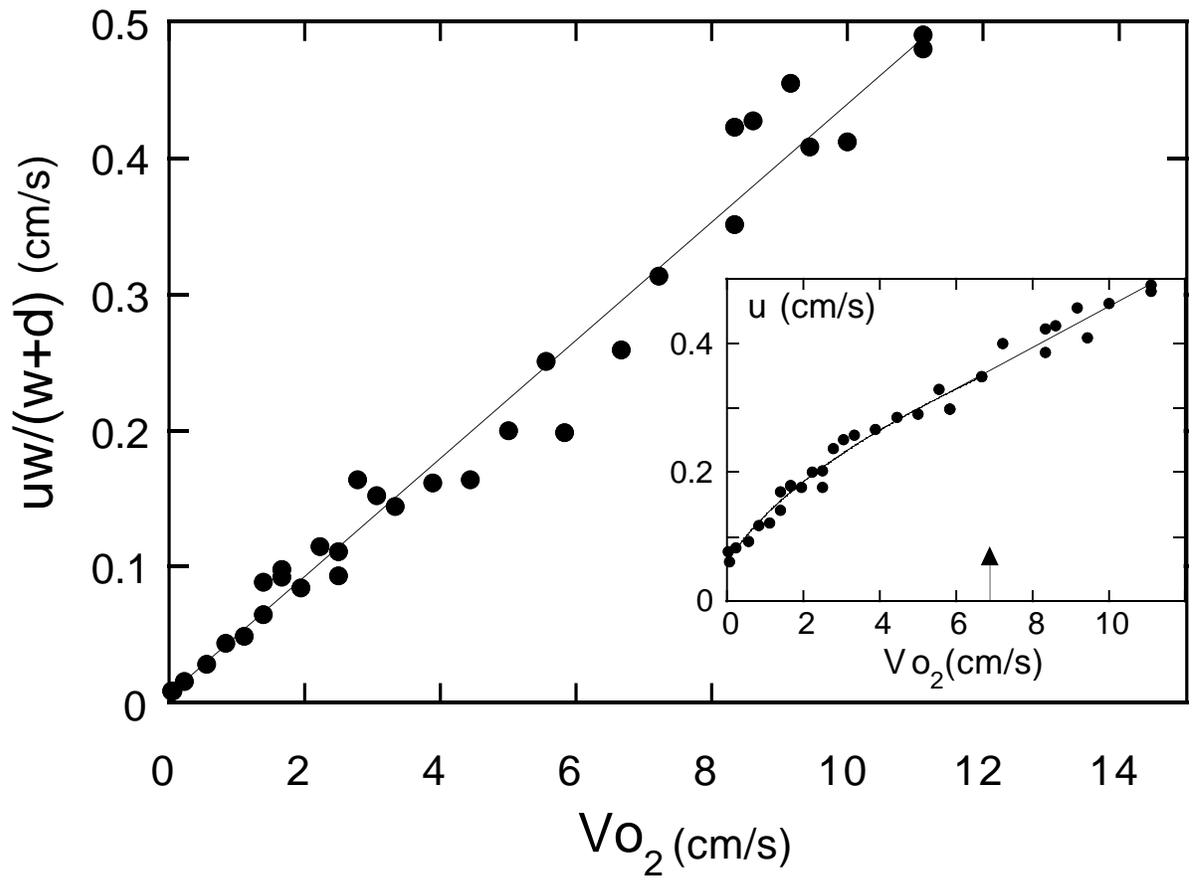
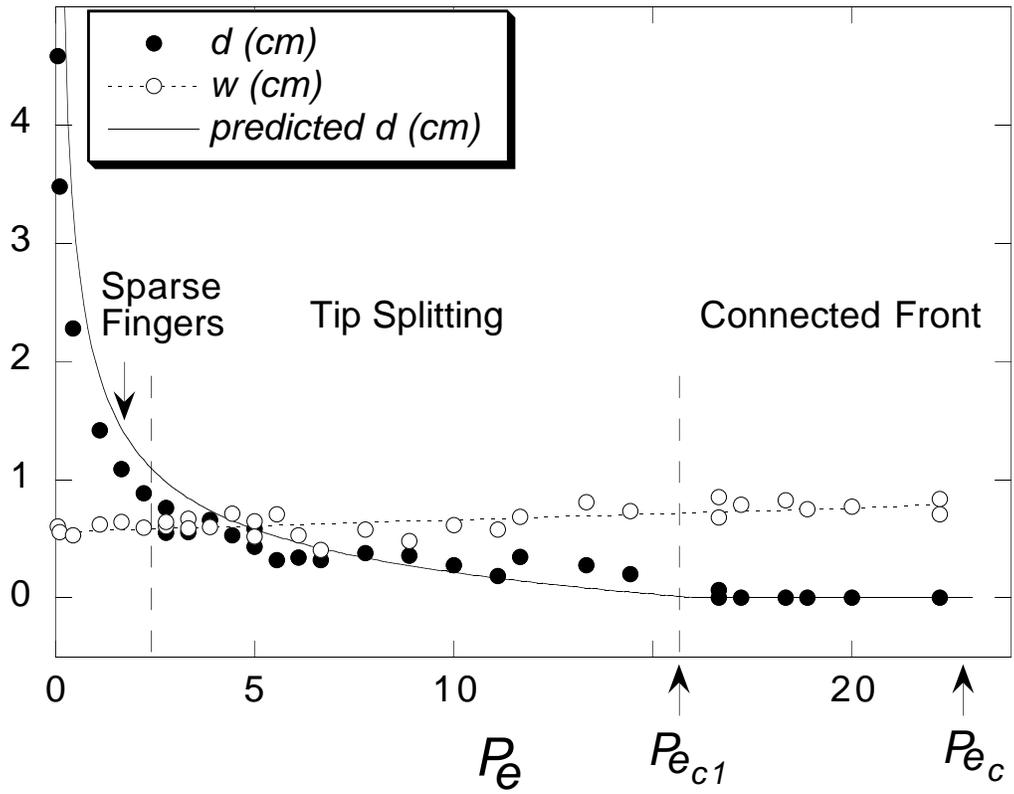
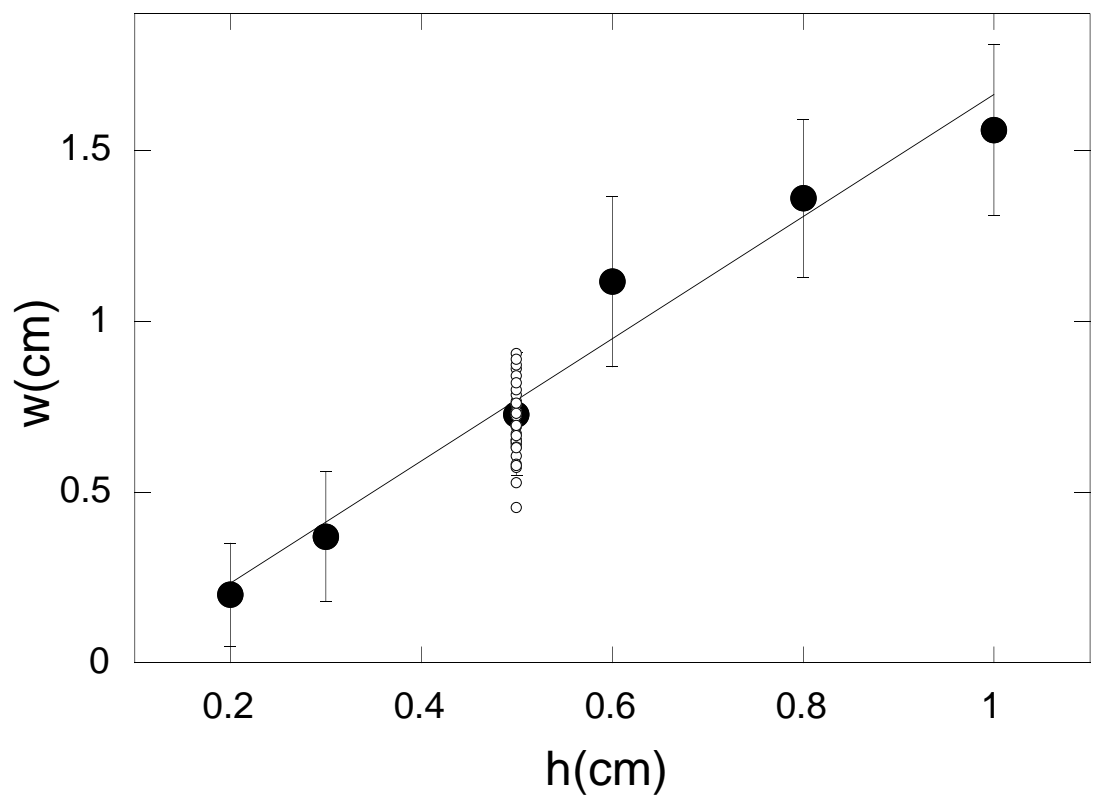
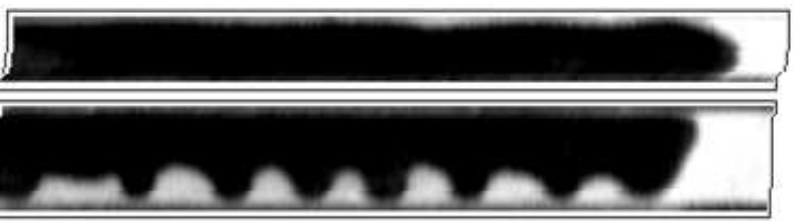
\end{document}